\documentstyle[preprint,aps,psfig]{revtex}
\newcommand{\beq}{\begin{equation}}
\newcommand{\eeq}{\end{equation}}
\newcommand{\beqa}{\begin{eqnarray}}
\newcommand{\eeqa}{\end{eqnarray}}
\newcommand{\ba}{\begin{array}}
\newcommand{\ea}{\end{array}}

\begin{document}
\draft 

\widetext 
\title{Formation of multi-solitons and 
vortex bright solitons in Bose-condensed alkali-metal atoms} 
\author{Luca Salasnich} 
\address{INFM-CNR, Unit\`a di Milano Universit\`a, \\
Dipartimento di Fisica, Universit\`a di Milano, \\ 
Via Celoria 16, Milano 20133, Italy \\
salasnich@mi.infm.it}
\maketitle 
\begin{abstract} 
The formation of multi-soliton configurations 
in Bose-condensed alkali-metal atoms 
is analyzed by using the nonpolynomial Schordinger equation. 
A train of bright solitons is obtained from an axially 
homogeneous Bose-Einstein condensate by a sudden change 
of the scattering length from repulsive to attractive.  
We derive an analytical expression for the number 
of bright solitons generated by using this mechanism. 
The formula generalizes a previous formula obtained 
with the 1D Gross-Pitaevskii equation. 
In the second part we consider vortex bright solitons, 
namely cigar-shaped bright solitons with a nonzero angular 
quantum number $k$ along the axial direction. By using a variational 
approach we determine the shape of vortex bright solitons, 
showing that the critical number of atoms for the collapse 
of the vortex soliton increases with a larger $k$. 
Finally we calculate monopole and quadrupole 
collective oscillations of these vortex bright solitons. 
\end{abstract} 

\vskip 1. truecm

\narrowtext

\newpage 

\section{Introduction}

In the last years we have investigated various dynamical properties 
of pure Bose-Einstein condensates (BECs) made of alkali-metal atoms 
at ultra-low temperatures \cite{1}. 
The recent experimental observation of dark \cite{2} and 
bright \cite{3,4} solitons in Bose-Einstein condensates (BECs) 
has focused our interest on BEC bright solitons. 
In particular, by using the nonpolynomial 
Schr\"odinger equation (NPSE), an effective equation 
derived from the 3D Gross-Pitaevskii 
equation (GPE) \cite{5}, we have studied the 
stability properties of bright solitons, 
determining the critical strength beyond 
which bright solitons collapse \cite{6}. 
This collapse is a direct consequence of the 
non-constant size of the soliton in the transverse 
direction, as found by solving the 3D GPE and NPSE. 
Instead, the 1D GPE, for which the transverse size of the 
soliton is constant, admits only stable bright 
solitons \cite{6}. 
\par 
In this paper we analyze the formation of bright solitons 
in a Bose-Einstein condensate of $^7$Li atoms 
induced by a sudden change in the sign 
of the scattering length from positive to negative, 
as reported in a recent experiment \cite{4}. 
In a previous paper \cite{7} we have show that a number of 
bright solitons is produced and this can be interpreted 
in terms of the modulational instability of the 
time-dependent macroscopic wave function of the Bose condensate. 
From the 1D GPE we have derived a simple formula for the number 
of solitons that is in good agreement with the numerical results. 
Here we generalize such a formula taking into account 
the non-constant size of the BEC in the transverse direction. 
Then we investigate the properties of vortex bright solitons, which 
are bright solitons with a finite quantum number of circulation 
around the longitudinal axis. 

\section{Formation of BEC bright solitons}

At zero temperature the macroscopic wave function of a BEC 
made of $N$ dilute $^7$Li 
atoms can be modeled by the 3D GPE. For a dilute gas the 
inter-atomic strength 
depend on the single parameter: the s-wave scattering length $a_s$. 
Note that only for $a_s<0$ there exist bright solitons, 
which are shape-invariant waves traveling without spreading 
due to the interplay between the repulsive kinetic energy 
and the attractive inter-atomic energy. 
\par 
Following the experiment of Strecker {\it et al.} \cite{4}, 
the trapping potential of the BEC is modeled by an anisotropic 
harmonic potential with transverse frequency 
$\omega_{\bot}= 2\pi \times 800$ Hz 
and longitudinal frequency $\omega_z=2\pi \times 4$ Hz. 
In addition there is a box optical potential that initially 
confines the condensate in the longitudinal direction. 
\par 
By using 3D GPE we first calculate the ground-state wave function 
of the condensate with a positive scattering length $a_s=100 a_B$, 
where $a_B=0.53\AA$ is the Bohr radius. 
We choose a condensate with longitudinal width 
$L=284.4$ $\mu$m and $N=10^4$ atoms. 
Then the ground-state wave function is inserted 
as initial condition for the time-evolution 
of the 3D GPE with $a_s=-3a_B$. 
We observe that a three-body dissipative 
term has been added to the 3D GPE. This 
term is important mainly during the collapse 
because it fixes the critical 
density at which the compression ceases (see also \cite{7}).  
\par 
In Figure 1 we plot the probability density 
in the longitudinal direction $\rho(z)$ of the evolving wave-function. 
The Bose-Einstein condensate shows the formation of 5 peaks. 
It is remarkable to stress that 
at the first instants the two external peaks 
become very high but the collapse is avoided by the 
dissipative term that quikly reduces their density. 
We have verified that without the dissipative term 
the Bose condensate collapses. The dissipative term 
does not alter the number of peaks but only their density. 
After the formation, the peaks start to separate each other showing a 
repulsive force between them. Eventually each peak evolves 
with small shape oscillations but without appreciable dispersion. 
\par 
The formation of these solitonic peaks can be explained 
as due to MI of the time-dependent wave function of the BEC, 
driven by imaginary Bogoliubov excitations \cite{7}. 
Neglecting the dissipative term, in the 1D limit of very long 
cigar-shaped condensate, the Bogoliubov longitudinal 
elementary excitations $\epsilon_k$ of the axially homogeneous 
BEC can be obtained from the 1D GPE. They are given by 
\beq 
\epsilon_k = \sqrt{ {\hbar^2 k^2\over 2m} 
\left({\hbar^2 k^2 \over 2 m} 
+ 2 g_{1d} n \right) }  , 
\eeq 
where $n=N/L$ is the linear axial density and 
$g_{1d}=2(\hbar\omega_{\bot})a_s$ 
is the 1D interaction strength.  
By suddenly changing the scattering length $a_s$ to a negative value, 
the excitations frequencies corresponding to 
$k<k_c = \left(16 \pi |g_{1d}| n\right)^{1/2}$ become 
imaginary and, as a result, small perturbations grow 
exponentially in time. 
It is easy to find that the maximum rate of growth is at 
$k_0=k_c/\sqrt{2}$. The wavelength of this mode is 
$\lambda_0 = {2\pi / k_0}$ and the ratio $L/\lambda_0$ gives 
an estimate of the number $N_s$ of bright solitons which 
are generated: 
\beq 
N_s = {\sqrt{N|a_s|L}\over \pi a_{\bot} } \; ,  
\eeq  
with $a_{\bot}=(\hbar/(m\omega_{\bot}))^{1/2}$. 
The predicted number $N_s$ of solitons 
is in very good agreement with the numerical results 
of Figure 1 (see also \cite{7}). 
\par 
The Eq. (2) is valid in the 1D limit 
but can be extended to a generic 3D case by using 
NPSE instead of 1D GPE. In general, 
the Bogoliubov excitations can be written as 
\beq  
\epsilon_k = \sqrt{ {\hbar^2 k^2\over 2m} 
\left({\hbar^2 k^2\over 2 m} + 2 c^2 \right) } \; , 
\eeq 
where $c$ is the first sound velocity in the longitudinal 
direction. The sound velocity $c$ satisfies the equation 
\beq 
c = \sqrt{n {\partial \mu \over \partial n} } \; , 
\eeq
where $\mu$ is the chemical potential of the 
Bose condensate. The chemical potential derived from 
1D GPE is $\mu = g_{1d} n +\hbar\omega_{\bot}$, 
while using NPSE one finds 
\beq 
\mu = { g_{1d} n \over \sqrt{1 + 2 a_s n} } 
+ {\hbar\omega_{\bot} \over 2} 
\left( {1 \over \sqrt{1+2 a_s n} } + 
\sqrt{1+2 a_s n} \right) \; . 
\eeq 
This chemical potential reduces to the 1D GPE one 
in the 1D limit, i.e. for $2a_s n <<1$. 
Moreover, if $a_s<0$ the chemical potential $\mu$ diverges 
at $2|a_s| n =1$, as a consequence of the 
collapse on the condensate \cite{8}. 
In conclusion, NPSE tells us that the number $N_s$ of bright 
solitons obtained via modulational instability is given by 
\beq 
N_s = {g \sqrt{2}\over 4 \pi} { \sqrt{4-3g\rho} 
\over \sqrt{g\rho} \left(1-g\rho\right)^{3/4}  } 
\; , 
\eeq 
where $g=2|a_s|/a_{\bot}$ is the adimensional inter-atomic 
strength and 
$\rho = n a_{\bot}$ is the adimensional axial density 
of the condensate. 
In Figure 2 we plot $N_s$ as a function of $g\rho$ for a fixed 
value of $\rho$. For small values of $g\rho$ 1D GPE and NPSE give 
the same results but approaching $g\rho =1$ the results are 
strongly different. In fact, NPSE gives a divergent number 
of solitons at $g\rho = 1$. As previously stated, only 
NPSE takes into account the non-constant width of the 
BEC in the transverse direction. Such a width shrinks  
to zero for a large attractive inter-atomic interaction. 

\section{Vortex bright solitons} 

The 3D GPE of a Bose-Einstein condensate of $N$ atoms 
confined in the transverse direction by a harmonic potential 
of frequency $\omega_{\bot}$ and with an attractive inter-atomic 
interaction ($a_s<0$) can be derived by the following scaled 
GPE Lagrangian 
\beq 
L = \psi^* \left[i {\partial \over \partial t} + {1\over 2}\nabla^2 
- {1\over 2}(x^2+y^2) - {2\pi \gamma } |\psi|^2 \right]\psi \; 
\eeq 
where $\psi$ is the macroscopic wave function of the condensate 
and $\gamma =Na_s/a_{\bot}$, 
with $a_{\bot}=(\hbar/(m\omega_{\bot}))^{1/2}$. 
Note that in the GPE Lagrangian lengths are in units $a_{\bot}$, 
time in units $\omega_{\bot}^{-1}$, 
action in units $\hbar$ and energy in units 
$\hbar \omega_{\bot}$. 
\par 
Using the cylindrical coordinates 
$r=(x^2+y^2)^{1/2}$, $\theta =arctg(y/x)$ and $z$, 
we write the trial wave function of the vortex bright soliton as 
\beq 
\psi (r,z,\theta,t) = r^k e^{ik\theta} 
\exp{ \left\{ 
-{r^2\over 2 \sigma(t)^2} 
-{z^2\over 2 \eta(t)^2} 
+i \alpha(t) r^2 + i \beta(t) z^2 
\right\} } \; , 
\eeq 
where $k$ is the vortex quantum number 
and $\sigma$, $\eta$, $\alpha$ and $\beta$ are 
the time-dependent variational parameters. 
In \cite{9} we have used this trial wave function 
to study vortex states 
with positive scattering length in a harmonic potential. 
Here we consider a negative scattering length. 
Moreover, in this case the (harmonic) confinement 
is only in the transverse direction. 
By inserting the trial wave function in the GPE Lagrangian, 
after spatial integration, one gets an effective Lagrangian depending 
on the variational parameters. By using the Euler-Lagrange equations 
one realizes that the time dependence of the phase parameters 
$\alpha$ and $\beta$ is fully determined by that of 
$\sigma$ and $\eta$ (see also \cite{9}). 
Thus, one obtains a new effective Lagrangian 
depending on $\sigma$ and $\eta$ only. It is given by 
\beq 
L_{eff} = (k+1){\dot \sigma}^2 + {1\over 2}{\dot \eta}^2 - 
(k+1) \sigma^2 - {(k+1)\over \sigma^2} - {1\over 2\eta^2} 
- \sqrt{2\over \pi}{(2k)! \over 2^{2k} (k!)^2} 
{\gamma \over \sigma^2\eta } \; . 
\eeq
In the case $k=0$ this effective Lagrangian reduces to the one 
we have successfully used to analyze the collective 
oscillations of the bright soliton in \cite{6}. 
In fact, the stationary solutions 
of the Euler-Lagrange equations of $L_{eff}$ 
give the widths of the bright soliton ($k=0$) and the widths 
of the vortex bright soliton ($k\neq 0$). 
\par 
In Figure 3 we plot the transverse radial width $\sigma$ and 
the longitudinal axial width $\eta$ of the vortex bright soliton 
as a function of the inter-atomic strength $\gamma$. 
For values of $\gamma$ close to zero the axial width $\eta$ 
is very large 
with respect to the radial width $\sigma$: the soliton is a strongly  
elongated cigar. By decreasing $\gamma$ the soliton anisotropy 
is reduced up to the collapse, i.e. the critical point at which  
the width parameters are no more minima of the GPE potential energy. 
It is interesting to observe that the critical $\gamma$ changes 
with the vortex quantum number $k$: $\gamma_c=-0.78$ for $k=0$, 
$\gamma_c=-2.20$ for $k=1$, and $\gamma_c=-3.59$ for $k=2$. 
This means that it is possible to avoid the collapse 
by creating vortex bright solitons with a large vortex 
quantum number $k$ (see also the recent 
numerical results \cite{10} of Adhikari). 
Another remarkable effect shown in Figure 3 for $k\neq 0$ 
is the crossing of the widths by decreasing $\gamma$: 
the vortex soliton becomes slightly disc-shaped. 
\par 
In the lower part of Figure 3 we plot the frequencies 
$\Omega_1$ and $\Omega_2$ of the collective oscillations   
of the vortex bright soliton along the axial and 
transverse directions. 
In cigar-shaped condensates $\Omega_1$ is called monopole 
frequency and $\Omega_2$ is called quadrupole frequency 
\cite{9}. These frequencies are obtained from 
the effective Lagrangian $L_{eff}$ linearizing 
around the minima of the potential energy 
(equilibrium solution). 
The transverse frequency $\Omega_2$ remains practically 
constant while the axial frequency $\Omega_1$ grows by 
decreasing $\gamma$ but goes to zero 
at the critical point, beyond which the frequencies 
are complex and the equilibrium solution becomes unstable. 
Note that for a fixed $\gamma$, 
the value of $\Omega_1$ strongly depends on $k$. 

\section{Conclusions}

In this paper we have shown that a train of bright solitons 
can be created in a Bose-Einstein condensate by a sudden 
change of the scattering length from positive to negative. 
The number of bright solitons can be predicted by using 
an analytical formula we have derived from the nonpolynomial 
Schr\"odinger equation. 
We have also investigated vortex bright solitons, which are 
bright solitons with a finite angular momentum along 
the longitudinal axis. Our results suggest that vortex 
bright solitons can be created 
with a larger number of atoms than bright solitons without 
vorticity. Moreover, our calculations suggest that 
vortex bright solitons can be experimentally detected 
by analyzing their collective oscillations.

\newpage

\section*{Figure captions}

~~~~
\vskip 0.5cm 

{\bf Figure 1}: Axial density profile $\rho(z)$ of the 
Bose-Einstein condensate made of $10^4$ $^7$Li atoms 
obtained by solving 3D GPE. 
For $t<0$ the scattering length is $a_s=100a_B$, while 
for $t\geq 0$ we set $a_s=-3a_B$ with $a_B$ the Bohr radius. 
Length $z$ in units $a_{z}=(\hbar /m\omega_z)^{1/2}$, 
density $\rho$ in units $1/a_z$ and 
time $t$ in units $\omega_z^{-1}$. 
Adapted from \cite{7}. 
\vskip 0.5cm 
{\bf Figure 2}: Number $N_s$ of bright solitons 
created via modulational instability. 
$\rho =Na_{\bot}/L$ is the initial (adimensional) axial density 
of the axially homogeneous Bose condensate. $g=2|a_s|/a_{\bot}$ is the 
inter-atomic strength with $a_{\bot}=(\hbar /m\omega_{\bot})^{1/2}$ 
and $a_s<0$ the final scattering length. 
\vskip 0.5cm 
{\bf Figure 3}: Top: transverse width $\sigma$ and axial width $\eta$ 
of the vortex bright soliton as a function of the interaction 
strength $\gamma=Na_s/a_{\bot}$. 
Bottom: monopole frequency $\Omega_1$ and quadrupole frequency 
$\Omega_2$ of the vortex bright soliton 
as a function of the interaction strength $\gamma$. 
Lengths in units $a_{\bot}$ and frequencies in units $\omega_{\bot}$. 
$k$ is the vortex quantum number. 


\newpage 

\begin{thebibliography}{99}

\bibitem{1} L. Salasnich, Int. J. Mod. Phys. B {\bf 15}, 1253 (2001);  
L. Salasnich, Int. J. Mod. Phys. B {\bf 16}, 2185 (2002); 
L. Salasnich, Nuovo Cimento B {\bf 114}, 637 (2002); 
L. Salasnich, Laser Phys. {\bf 13}, 547 (2003); 

\bibitem{2} S. Burger {et al.}, Phys. Rev. lett. {\bf 83}, 5198 (1999);   
J. Denshlag {\it et al.}, Science {\bf 287}, 97 (2000). 

\bibitem{3} L. Khaykovich {\it et al.}, Science {\bf 296}, 1290 (2002). 

\bibitem{4} K.E. Strecker {\it et al.}, 
Nature {\bf 417}, 150 (2002). 

\bibitem{5} L. Salasnich, Laser Phys. {\bf 12}, 198 (2002); 
L. Salasnich, A. Parola and L. Reatto, Phys. Rev. A {\bf 65}, 043614 (2002); 
L. Salasnich, A. Parola and L. Reatto, 
J. Phys. B (Atom. Mol.) {\bf 35}, 3205 (2002). 
L. Salasnich, Laser Phys. {\bf 13}, 543 (2003).   
 
\bibitem{6} L. Salasnich, A. Parola and L. Reatto,   
Phys. Rev. A {\bf 66}, 043603 (2002); 
L. Salasnich, Progr. Theor. Phys. Suppl. {\bf 150} (2003). 

\bibitem{7} L. Salasnich, A. Parola and L. Reatto,   
cond-mat/0307206, to be published in Phys. Rev. Lett. (2003).  

\bibitem{8} L. Salasnich, A. Parola and L. Reatto, 
in preparation. 

\bibitem{9} L. Salasnich, Int. J. Mod. Phys. B {\bf 14}, 1 (2000). 

\bibitem{10} S.K. Adhikari, Phys. Rev. E {\bf 65}, 016703 (2002). 

\end{thebibliography}
\end{document}